\documentclass[12pt,a4paper]{article}

\usepackage{ifthen} 
\newboolean{pdflatex}
\setboolean{pdflatex}{true} 

\newboolean{articletitles}
\setboolean{articletitles}{true} 

\newboolean{uprightparticles}
\setboolean{uprightparticles}{false} 

\newboolean{inbibliography}
\setboolean{inbibliography}{false} 

\usepackage{microtype}
\usepackage{lineno}  
\usepackage{xspace} 
\usepackage{caption} 

\usepackage{graphicx}  
\usepackage{color}
\usepackage{colortbl}
\graphicspath{{./figs/}} 

\usepackage{amsmath} 
\usepackage{amssymb}
\usepackage{amsfonts}
\usepackage{upgreek} 

\newcommand*\patchAmsMathEnvironmentForLineno[1]{%
\expandafter\let\csname old#1\expandafter\endcsname\csname #1\endcsname
\expandafter\let\csname oldend#1\expandafter\endcsname\csname
end#1\endcsname
 \renewenvironment{#1}%
   {\linenomath\csname old#1\endcsname}%
   {\csname oldend#1\endcsname\endlinenomath}%
}
\newcommand*\patchBothAmsMathEnvironmentsForLineno[1]{%
  \patchAmsMathEnvironmentForLineno{#1}%
  \patchAmsMathEnvironmentForLineno{#1*}%
}
\AtBeginDocument{%
\patchBothAmsMathEnvironmentsForLineno{equation}%
\patchBothAmsMathEnvironmentsForLineno{align}%
\patchBothAmsMathEnvironmentsForLineno{flalign}%
\patchBothAmsMathEnvironmentsForLineno{alignat}%
\patchBothAmsMathEnvironmentsForLineno{gather}%
\patchBothAmsMathEnvironmentsForLineno{multline}%
}

\usepackage{hyperref}    
\usepackage[all]{hypcap} 

\def\lhcb {\mbox{LHCb}\xspace}

\def\velo   {VELO\xspace}

\ifthenelse{\boolean{uprightparticles}}%
{

 \def\Ppi         {\ensuremath{\uppi}\xspace}

 \def\PDelta      {\ensuremath{\Delta}\xspace}                 
 \def\PXi      {\ensuremath{\Xi}\xspace}                 
 \def\PLambda      {\ensuremath{\Lambda}\xspace}                 
 \def\PSigma      {\ensuremath{\Sigma}\xspace}                 
 \def\POmega      {\ensuremath{\Omega}\xspace}                 
 \def\PUpsilon      {\ensuremath{\Upsilon}\xspace}                 
 
 \def\PB      {\ensuremath{\mathrm{B}}\xspace}                 
                  
 \def\PD      {\ensuremath{\mathrm{D}}\xspace}

 \def\PK      {\ensuremath{\mathrm{K}}\xspace}

 \def\Pb      {\ensuremath{\mathrm{b}}\xspace}                 
 \def\Pc      {\ensuremath{\mathrm{c}}\xspace}

 \def\Ph      {\ensuremath{\mathrm{h}}\xspace}                 
 \def\Pi      {\ensuremath{\mathrm{i}}\xspace}

 \def\Ps      {\ensuremath{\mathrm{s}}\xspace}

}
{

 \def\Ppi         {\ensuremath{\pi}\xspace}

 \mathchardef\PDelta="7101
 \mathchardef\PXi="7104
 \mathchardef\PLambda="7103
 \mathchardef\PSigma="7106
 \mathchardef\POmega="710A
 \mathchardef\PUpsilon="7107
                  
 \def\PB      {\ensuremath{B}\xspace}                 
                  
 \def\PD      {\ensuremath{D}\xspace}

 \def\PK      {\ensuremath{K}\xspace}

 \def\Pb      {\ensuremath{b}\xspace}                 
 \def\Pc      {\ensuremath{c}\xspace}

 \def\Ph      {\ensuremath{h}\xspace}                 
 \def\Pi      {\ensuremath{i}\xspace}

 \def\Ps      {\ensuremath{s}\xspace}

}

\def\squark    {{\ensuremath{\Ps}}\xspace}

\def\cquark    {{\ensuremath{\Pc}}\xspace}

\def\bquark    {{\ensuremath{\Pb}}\xspace}

\def\pion   {{\ensuremath{\Ppi}}\xspace}
\def\piz    {{\ensuremath{\pion^0}}\xspace}

\def\pip    {{\ensuremath{\pion^+}}\xspace}
\def\pim    {{\ensuremath{\pion^-}}\xspace}

\def\kaon    {{\ensuremath{\PK}}\xspace}
  \def\Kbar    {{\kern 0.2em\overline{\kern -0.2em \PK}{}}\xspace}

\def\Kp      {{\ensuremath{\kaon^+}}\xspace}
\def\Km      {{\ensuremath{\kaon^-}}\xspace}

\def\KS      {{\ensuremath{\kaon^0_{\rm\scriptscriptstyle S}}}\xspace}

  \def\Dbar    {{\kern 0.2em\overline{\kern -0.2em \PD}{}}\xspace}
\def\D       {{\ensuremath{\PD}}\xspace}

\def\Dz      {{\ensuremath{\D^0}}\xspace}

\def\B       {{\ensuremath{\PB}}\xspace}
\def\Bbar    {{\ensuremath{\kern 0.18em\overline{\kern -0.18em \PB}{}}}\xspace}

\def\Bd      {{\ensuremath{\B^0}}\xspace}
\def\Bs      {{\ensuremath{\B^0_\squark}}\xspace}
\def\Bsb     {{\ensuremath{\Bbar^0_\squark}}\xspace}

\def\Bc      {{\ensuremath{\B_\cquark^+}}\xspace}

\def\Bbracks      {{\ensuremath{\B^0_{(\squark)}}}\xspace}
\def\Bbracksb     {{\ensuremath{\Bbar^0_{(\squark)}}}\xspace}

  \def\Y#1S{\ensuremath{\PUpsilon{(#1S)}}\xspace}

\def\Lbar        {{\ensuremath{\kern 0.1em\overline{\kern -0.1em\PLambda}}}\xspace}

\newcommand{\decay}[2]{\ensuremath{#1\!\to #2}\xspace}         
\def\ra                 {\ensuremath{\rightarrow}\xspace}
\def\to                 {\ensuremath{\rightarrow}\xspace}

\def\CP                {{\ensuremath{C\!P}}\xspace}

\newcommand{\DG}{{\ensuremath{\Delta\Gamma}}\xspace}
\newcommand{\DGs}{{\ensuremath{\Delta\Gamma_{\squark}}}\xspace}

\newcommand{\Gs}{{\ensuremath{\Gamma_{\squark}}}\xspace}

\newcommand{\ADGs}{\ensuremath{{\cal A}_{\Delta\Gamma_\squark}}\xspace}

\newcommand{\ADGbracks}{\ensuremath{{\cal A}_{\Delta\Gamma_{(\squark)}}}\xspace}

\newcommand{\DGbracks}{{\ensuremath{\Delta\Gamma_{(\squark)}}}\xspace}

\def\BTohh        {\decay{\B^{0}_{(s)}}{\Ph^+ \Ph'^-}}

\def\BdToKpi      {\decay{\Bd}{\Kp\pim}}
\def\BsToKK       {\decay{\Bs}{\Kp\Km}}
\def\BsTopiK      {\decay{\Bs}{\pip\Km}}
\def\BdToKK      {\decay{\Bd}{\Kp\Km}}

\def\AT#1     {\ensuremath{A_{\mathrm{T}}^{#1}}\xspace}           

\def\C#1      {\ensuremath{\mathcal{C}_{#1}}\xspace}                       
\def\Cp#1     {\ensuremath{\mathcal{C}_{#1}^{'}}\xspace}                    
\def\Ceff#1   {\ensuremath{\mathcal{C}_{#1}^{\mathrm{(eff)}}}\xspace}        
\def\Cpeff#1  {\ensuremath{\mathcal{C}_{#1}^{'\mathrm{(eff)}}}\xspace}       
\def\Ope#1    {\ensuremath{\mathcal{O}_{#1}}\xspace}                       
\def\Opep#1   {\ensuremath{\mathcal{O}_{#1}^{'}}\xspace}                    

\newcommand{\tev}{\ifthenelse{\boolean{inbibliography}}{\ensuremath{~T\kern -0.05em eV}\xspace}{\ensuremath{\mathrm{\,Te\kern -0.1em V}}}\xspace}
\newcommand{\gev}{\ensuremath{\mathrm{\,Ge\kern -0.1em V}}\xspace}
\newcommand{\mev}{\ensuremath{\mathrm{\,Me\kern -0.1em V}}\xspace}
\newcommand{\kev}{\ensuremath{\mathrm{\,ke\kern -0.1em V}}\xspace}
\newcommand{\ev}{\ensuremath{\mathrm{\,e\kern -0.1em V}}\xspace}
\newcommand{\gevc}{\ensuremath{{\mathrm{\,Ge\kern -0.1em V\!/}c}}\xspace}
\newcommand{\mevc}{\ensuremath{{\mathrm{\,Me\kern -0.1em V\!/}c}}\xspace}
\newcommand{\gevcc}{\ensuremath{{\mathrm{\,Ge\kern -0.1em V\!/}c^2}}\xspace}
\newcommand{\gevgevcccc}{\ensuremath{{\mathrm{\,Ge\kern -0.1em V^2\!/}c^4}}\xspace}
\newcommand{\mevcc}{\ensuremath{{\mathrm{\,Me\kern -0.1em V\!/}c^2}}\xspace}

\def\invpb {\ensuremath{\mbox{\,pb}^{-1}}\xspace}

\def\invfb   {\ensuremath{\mbox{\,fb}^{-1}}\xspace}

\def\ps   {\ensuremath{{\rm \,ps}}\xspace}

\newcommand{\stat}{\ensuremath{\mathrm{\,(stat)}}\xspace}
\newcommand{\syst}{\ensuremath{\mathrm{\,(syst)}}\xspace}

\newcommand{\chisqndf}{\ensuremath{\chi^2/\mathrm{ndf}}\xspace}
\newcommand{\chisqip}{\ensuremath{\chi^2_{\rm IP}}\xspace}

\def\gsim{{~\raise.15em\hbox{$>$}\kern-.85em
          \lower.35em\hbox{$\sim$}~}\xspace}
\def\lsim{{~\raise.15em\hbox{$<$}\kern-.85em
          \lower.35em\hbox{$\sim$}~}\xspace}

\def\sPlot{\mbox{\em sPlot}}

\def\sWeight{\mbox{\em sWeight}}
\def\sWeights{\mbox{\em sWeights}}

\def\pt         {\mbox{$p_{\rm T}$}\xspace}

\def\mrad{\ensuremath{\rm \,mrad}\xspace}

\def\evtgen     {\mbox{\textsc{EvtGen}}\xspace}

\def\geant      {\mbox{\textsc{Geant4}}\xspace}

\def\photos     {\mbox{\textsc{Photos}}\xspace}

\def\pythia     {\mbox{\textsc{Pythia}}\xspace}

\def\tell1  {TELL1\xspace}
\def\ukl1   {UKL1\xspace}

\usepackage{cite} 
\usepackage{mciteplus}

\usepackage{longtable} 

\begin{document}

\renewcommand{\thefootnote}{\fnsymbol{footnote}}
\setcounter{footnote}{1}


\begin{titlepage}
\pagenumbering{roman}

\vspace*{-1.5cm}
\centerline{\large EUROPEAN ORGANIZATION FOR NUCLEAR RESEARCH (CERN)}
\vspace*{1.5cm}
\hspace*{-0.5cm}
\begin{tabular*}{\linewidth}{lc@{\extracolsep{\fill}}r}
\ifthenelse{\boolean{pdflatex}}
{\vspace*{-2.7cm}\mbox{\!\!\!\includegraphics[width=.14\textwidth]{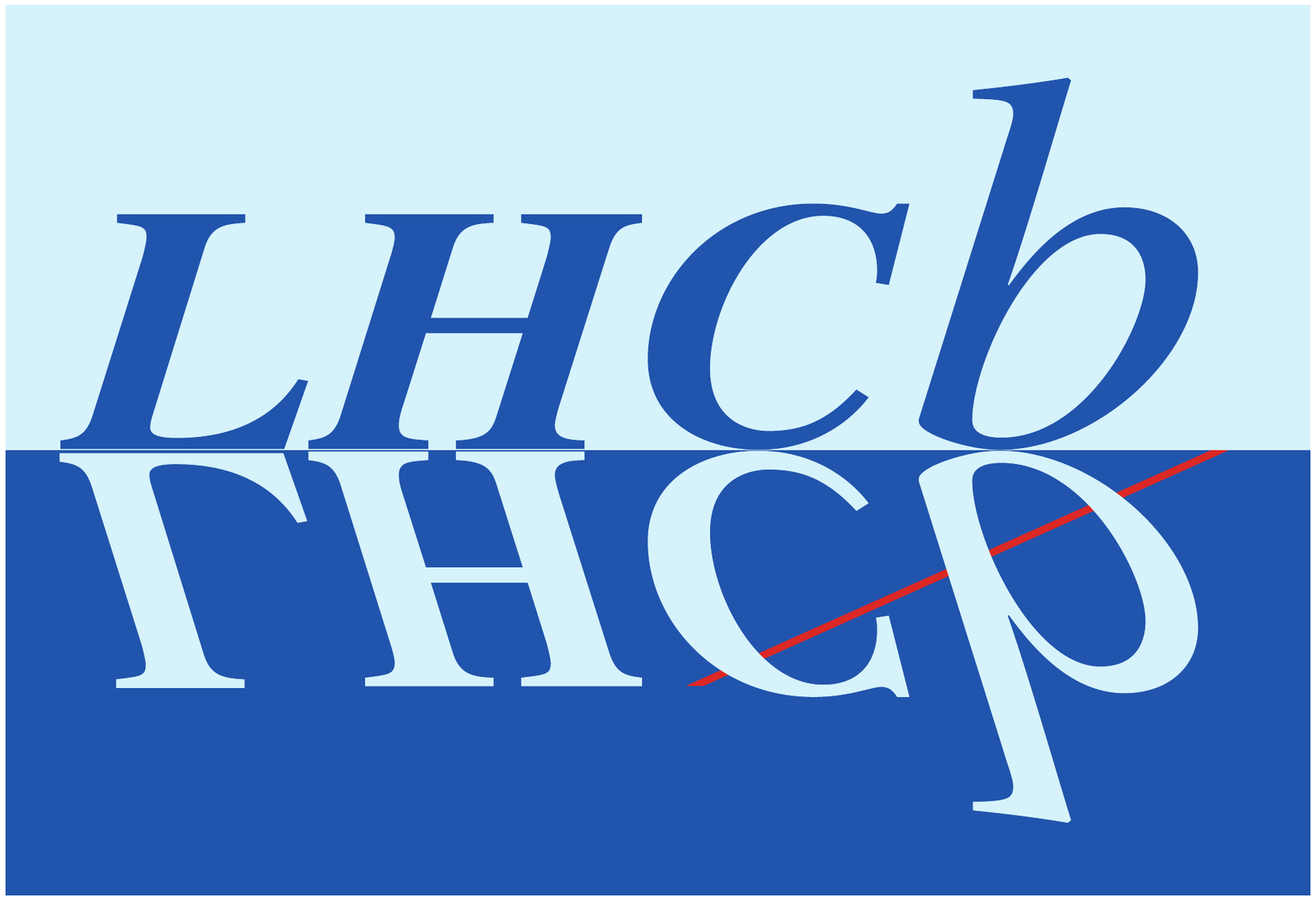}} & &}%
{\vspace*{-1.2cm}\mbox{\!\!\!\includegraphics[width=.12\textwidth]{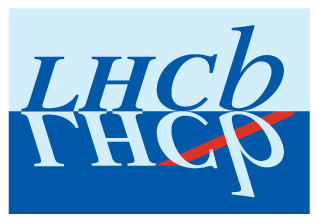}} & &}%
\\
 & & CERN-PH-EP-2014-127 \\  
 & & LHCb-PAPER-2014-011 \\  
 & & \today \\ 
 & & \\
\end{tabular*}

\vspace*{2.0cm}

{\bf\boldmath\huge
\begin{center}
  Effective lifetime measurements in the \BsToKK, \BdToKpi and \BsTopiK decays 
\end{center}
}

\vspace*{2.0cm}

\begin{center}
The LHCb collaboration\footnote{Authors are listed on the following pages.}
\end{center}

\vspace{\fill}

\begin{abstract}
 
\noindent Measurements of the effective lifetimes in the \BsToKK, \BdToKpi and \BsTopiK 
decays are presented using $1.0\invfb$ of $pp$ collision data collected at a centre-of-mass energy of $7\tev$ by the \lhcb
experiment. 
The analysis uses a data-driven approach to correct
for the decay time acceptance.
The measured effective 
lifetimes are
\begin{eqnarray*}
  \tau_{\BsToKK} &=& 1.407~\pm~0.016~\stat~\pm~0.007~\syst~\ps, \\
  \tau_{\BdToKpi}     &=& 1.524~\pm~0.011~\stat~\pm~0.004~\syst~\ps, \\
  \tau_{\BsTopiK}     &=& 1.60~\pm~0.06~\stat~\pm~0.01~\syst~\ps. 
\end{eqnarray*}

\noindent 
This is the most precise determination to date of the effective lifetime in the 
\BsToKK decay and provides
constraints on contributions from physics beyond the Standard Model to the \Bs mixing phase and
the width difference \DGs. 
  
\end{abstract}

\vspace*{1.0cm}

\begin{center}
  Submitted to Phys.~Lett.~B 
\end{center}

\vspace{\fill}

{\footnotesize 
\centerline{\copyright~CERN on behalf of the \lhcb collaboration, license \href{http://creativecommons.org/licenses/by/3.0/}{CC-BY-3.0}.}}
\vspace*{2mm}

\end{titlepage}


\newpage
\setcounter{page}{2}
\mbox{~}
\newpage


\centerline{\large\bf LHCb collaboration}
\begin{flushleft}
\small
R.~Aaij$^{41}$, 
B.~Adeva$^{37}$, 
M.~Adinolfi$^{46}$, 
A.~Affolder$^{52}$, 
Z.~Ajaltouni$^{5}$, 
J.~Albrecht$^{9}$, 
F.~Alessio$^{38}$, 
M.~Alexander$^{51}$, 
S.~Ali$^{41}$, 
G.~Alkhazov$^{30}$
% ...

{\footnotesize \it
% ...
$ ^{5}$Clermont Universit\'{e}, Universit\'{e} Blaise Pascal, CNRS/IN2P3, LPC, Clermont-Ferrand, France\\
% ...
$ ^{9}$Fakult\"{a}t Physik, Technische Universit\"{a}t Dortmund, Dortmund, Germany\\
% ...
$ ^{30}$Petersburg Nuclear Physics Institute (PNPI), Gatchina, Russia\\
% ...
$ ^{37}$Universidad de Santiago de Compostela, Santiago de Compostela, Spain\\
$ ^{38}$European Organization for Nuclear Research (CERN), Geneva, Switzerland\\
% ...
$ ^{41}$Nikhef National Institute for Subatomic Physics, Amsterdam, The Netherlands\\
% ...
$ ^{46}$H.H. Wills Physics Laboratory, University of Bristol, Bristol, United Kingdom\\
% ...
$ ^{51}$School of Physics and Astronomy, University of Glasgow, Glasgow, United Kingdom\\
$ ^{52}$Oliver Lodge Laboratory, University of Liverpool, Liverpool, United Kingdom\\
% ...
}
\end{flushleft}



\cleardoublepage


\renewcommand{\thefootnote}{\arabic{footnote}}
\setcounter{footnote}{0}



\pagestyle{plain} 
\setcounter{page}{1}
\pagenumbering{arabic}


%

\section{Introduction}
\label{sec:Introduction}

The study of $\PB^{0}_{(s)}$ mesons from the charmless $\BTohh$ decay family\footnote{The inclusion of charge-conjugate processes is implied.}, where $h^{(\prime)}$ is either a pion or a kaon, offers unique opportunities to investigate the heavy flavour sector. These decays are sensitive to charge parity (\CP) symmetry violation, which allows the phase structure of the Cabibbo-Kobayashi-Maskawa (CKM) matrix~\cite{bib:Cabibbo, bib:CKM1973} to be studied, and to manifestations of physics beyond the Standard Model (SM).  
The \BTohh decays have been analysed in detail by \lhcb, with measurements of the branching fractions~\cite{bib:B2hh_2012as}, time-integrated~\cite{bib:lhcbB2hhTICP} and time-dependent~\cite{bib:ProductionAsymmetry2013} \CP violation being made.  
The effective \BsToKK lifetime has previously been measured by \lhcb using data recorded in 2010~\cite{bib:Glasgow_LHCb_BsKKLifetime_2010_Paper} and 2011~\cite{bib:lhcbBs2KKUnbiased}, corresponding to an integrated luminosity of $37\invpb$ and $1.0\invfb$ respectively. In this paper we reanalyse the 2011 data using a data driven method that employs the full statistical power of the data set.

The detailed formalism of the effective lifetime in \BTohh decays can be found in Refs.~\cite{bib:Fleischer1} and ~\cite{bib:Fleischer2}.  The decay time distribution of a \BTohh decay, with equal contributions of both $B^{0}_{(s)}$ and $\bar{B}^{0}_{(s)}$ at the production stage, can be written as 

\begin{eqnarray}
\label{eq:untaggedLTEqn}
  \Gamma(t) & \propto & \left ( 1 - \ADGbracks  \right )e^{-\Gamma_{\mathrm{L}}^{(s)} t} + \left ( 1 + \ADGbracks \right )e^{-\Gamma_{\mathrm{H}}^{(s)} t} \, ,
\end{eqnarray}

\noindent where $\Gamma_{\mathrm{H}}^{(s)} = \Gamma_{(s)} - \DGbracks/2$ and $\Gamma_{\mathrm{L}}^{(s)} =  \Gamma_{(s)} + \DGbracks/2$ are the decay widths of the heavy and light mass eigenstates, $\Gamma_{(s)}$ is the average decay width and $\DG_{(s)}$ is the decay width difference between the mass eigenstates. These in turn are given as linear combinations of the two flavour states with complex coefficients $q$ and $p$. The formalism used herein is only valid if $|q/p|=1$.

The parameter \ADGbracks is defined as $\ADGbracks \equiv -2 {\rm Re}(\lambda)/\left(1 + |\lambda|^2\right)$, where $\lambda \equiv (q/p)(\overline{A}/A)$ and 
 $A$ ($\overline{A}$) is the amplitude for \Bbracks (\Bbracksb) decays to the respective final states.
For \Bd mesons, \DG is sufficiently small that the heavy and light mass eigenstates cannot be resolved experimentally, thus only a single exponential distribution is measured. 
For \Bs mesons, \DGs is large enough for the mass eigenstates to be distinguishable. 
This implies that fitting a single exponential distribution will yield a different \textit{effective} lifetime when measured in different \Bs channels, depending on the relative proportions of the heavy and light contributions in that decay.
Equal proportions of heavy and light eigenstates contribute to the \BsTopiK decay at $t=0$, which allows measuring the flavour-specific effective lifetime. The \Bs flavour-specific effective lifetime can be approximated to second order by

\begin{equation}
\label{eq:tauBsEffLifetime}
\tau_{\Bs} \approx \frac{1}{\Gs} \frac{1 + \left( \frac{\DGs}{2\Gs} \right)^2}{1 - \left( \frac{\DGs}{2\Gs} \right)^2}~.
\end{equation}

The \BsToKK decay is treated slightly differently as the SM predicts the initial state to consist almost entirely of the light mass eigenstate. This can be described by stating that in the absence of \CP violation the parameter $\ADGs(\BsToKK) = -1$, thus the decay time distribution involves only the first term in Eq.~\ref{eq:untaggedLTEqn}. 
For small deviations from the \CP-conserving limit, the distribution can be approximated to first order in \DGs/\Gs by a single exponential with an effective lifetime

\begin{eqnarray}
\label{eq:tauKKEqn}
  \tau_{\BsToKK} & \approx & \frac{1}{\Gs} \left( 1 + \frac{\ADGs \DGs}{2\Gs} \right) \,. 
\end{eqnarray}

\noindent An effective lifetime measurement in the decay channel \BsToKK is of considerable interest, as it can be used to constrain the contributions from new physical phenomena entering the \Bs meson system~\cite{bib:Grossman,Lenz:2006hd,bib:Fleischer1,bib:Fleischer2,bib:Fleischer_2011cw}. This decay channel has contributions from loop diagrams that in the SM have the same phase as the $\Bs$--$\Bsb$ mixing amplitude, hence the measured effective lifetime is expected to be close to $1/\Gamma_{\mathrm{L}}^{s}$. However, the tree contribution to the \BsToKK decay amplitude introduces a small amount of \CP violation.   
Taking the SM prediction for $\ADGs(\BsToKK) = -0.972^{+0.014}_{-0.009}$~\cite{bib:Fleischer1} and 
the measured values of \Gs and \DGs from Ref.~\cite{bib:LHCB-PAPER-2013-002}, the prediction for the effective \BsToKK lifetime from Eq.~\ref{eq:tauKKEqn} is $\tau_{\BsToKK} = 1.395\pm0.020\ps$.

The measurement is performed using a $pp$ collision data sample corresponding to an integrated luminosity of $1.0\invfb$, collected by the \lhcb experiment at a centre of mass energy of $\sqrt{s} = 7\tev$  in 2011.  
A key aspect of the analysis is the correction of decay time biasing effects, referred to as the acceptance, which are introduced by the selection criteria used to maximise the signal significance of the \B meson sample.  A data-driven approach, discussed in detail in Ref.~\cite{bib:AGamma2011}, and applied to a previous measurement of this channel~\cite{bib:Glasgow_LHCb_BsKKLifetime_2010_Paper}, is used to correct for this bias.  

\section{Detector and data sample}
The \lhcb detector~\cite{bib:LHCb} is a single-arm forward
spectrometer covering the \mbox{pseudorapidity} range $2<\eta <5$,
designed for the study of particles containing \bquark or \cquark
quarks. The detector includes a high-precision tracking system 
consisting of a silicon-strip vertex detector (\velo) surrounding the $pp$ interaction region~\cite{bib:LHCb-DP-2014-001}
and several dedicated tracking planes
with silicon microstrip detectors (Inner Tracker) covering the region
with high charged particle multiplicity and straw tube detectors
(Outer Tracker) for the region with lower occupancy. The Inner and
Outer Tracker are placed 
downstream of the magnets to allow the measurement
of the charged particles momenta as they traverse the detector.
Excellent particle identification (PID) capabilities are provided by two
ring-imaging Cherenkov detectors, which allow charged pions, kaons, and
protons to be distinguished from each other in the momentum range
2--100\gevc~\cite{bib:RICHPerformance2012}. The experiment employs a multi-level trigger to reduce
the readout rate and enhance signal purity: a hardware trigger based
on the measurement of the transverse energy deposited in the calorimeter cells
and the momentum transverse to the beamline (\pt) of muon candidates, as
well as a software trigger that allows the reconstruction of the full
event information.

The average momentum of the produced \B mesons is around 100\gevc
and their decay vertices are displaced from the primary interaction
vertex (PV). Background particles in general have low momentum and
originate from the primary $pp$ collision. 
The candidates used in this analysis are reconstructed from events selected by the hardware trigger and containing 
large hadronic energy depositions that originate from the signal particles, 
or events selected that do not originate from the signal particles.
The signal sample is further enriched by the software-based trigger with an exclusive selection on \BTohh candidates. 

The offline selection is based on a cut-based method, which is designed
to maximise the signal significance. The selection requires that the tracks associated with the
\B meson decay products
have a good track fit quality per number of degrees of freedom, $\chisqndf< 3.3$.
The transverse momentum of at least one particle from the decay is required to have 
$\pt > 2.5 \gevc$, with
the other having $\pt > 1.1 \gevc$. Each decay product must also have a large \chisqip,  
defined as the difference in $\chi^{2}$ of the primary $pp$ interaction vertex reconstructed with and without the considered particle.
The minimum value of the \chisqip of the two decay products is required to be greater than $45$, and the larger of the two greater than $70$. 

The \B meson candidate is obtained by reconstructing the vertex formed
by the two particles. It is required to have 
$\mathrm{\chisqip < 9}$ and a reconstructed decay time greater than $0.6 \ps$. 
Each $pp$ interaction vertex  in an event is fitted with both the reconstructed charged particles, 
where there are are typically $1.7$ interaction vertices per bunch crossing.
The angle between the direction of flight from the best PV to decay vertex, and
the \B momentum vector, must be smaller than $19 \mrad$. The best PV is defined as 
the PV to which the \B candidate has the lowest \chisqip value.

The final selection of the \BTohh modes is performed by identifying
pions, kaons and protons using 
PID likelihood observables 
obtained from the ring-imaging Cherenkov
detectors~\cite{bib:RICHPerformance2012}.
Simulated samples of these \BTohh modes are also generated for verification.
In the simulation, $pp$ collisions are generated using
\pythia~\cite{Sjostrand:2006za,*Sjostrand:2007gs} 
with a specific \lhcb
configuration~\cite{LHCb-PROC-2010-056}.  Decays of hadronic particles
are described by \evtgen~\cite{Lange:2001uf}, in which final state
radiation is generated using \photos~\cite{Golonka:2005pn}. The
interaction of the generated particles with the detector and its
response are implemented using the \geant
toolkit~\cite{Allison:2006ve, *Agostinelli:2002hh} as described in
Ref.~\cite{LHCb-PROC-2011-006}.

\section{\BTohh lifetime measurements}

\begin{figure}[htbp]
  \begin{center}
     
     \includegraphics*[width=0.49\textwidth]{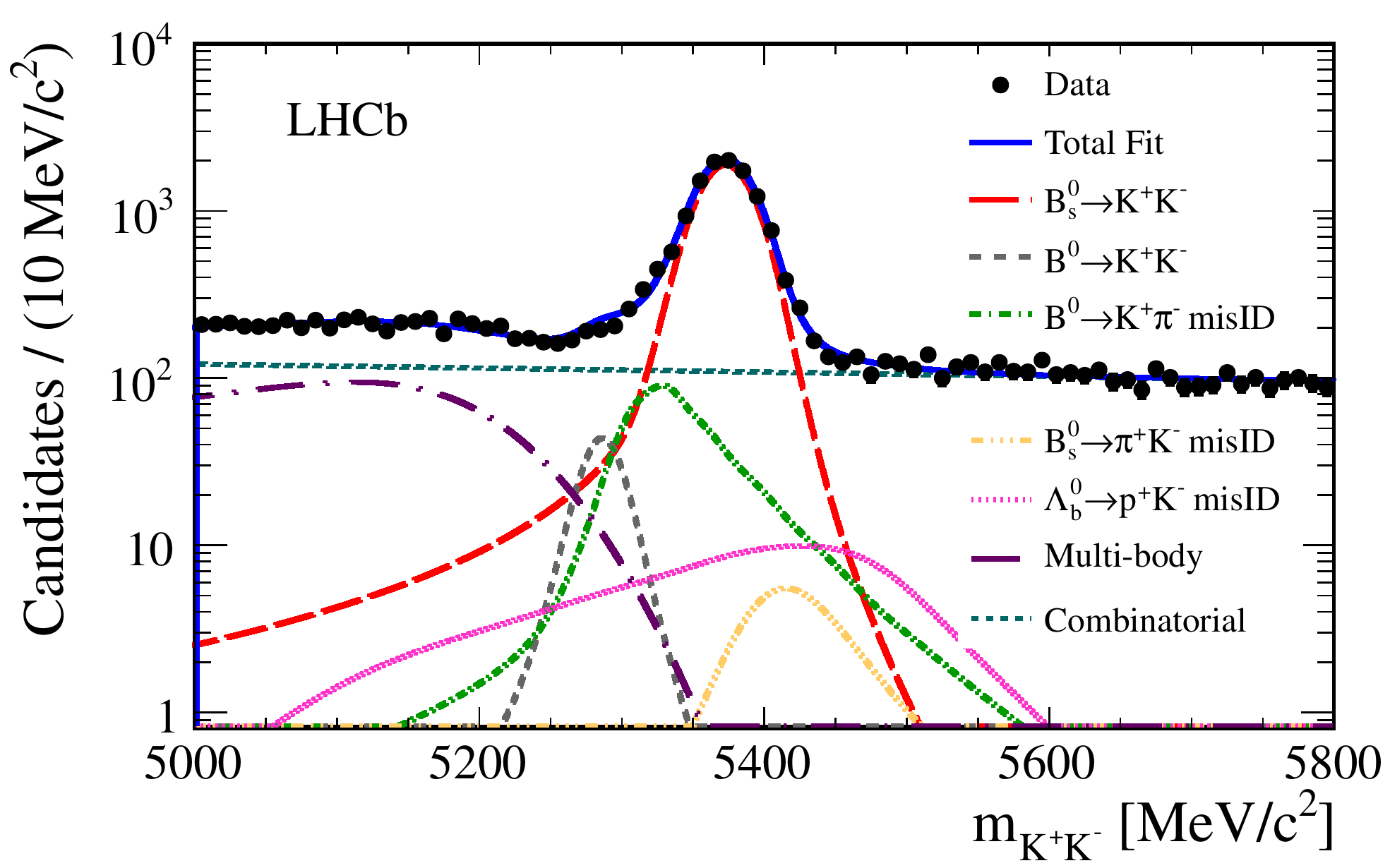}
     \includegraphics*[width=0.49\textwidth]{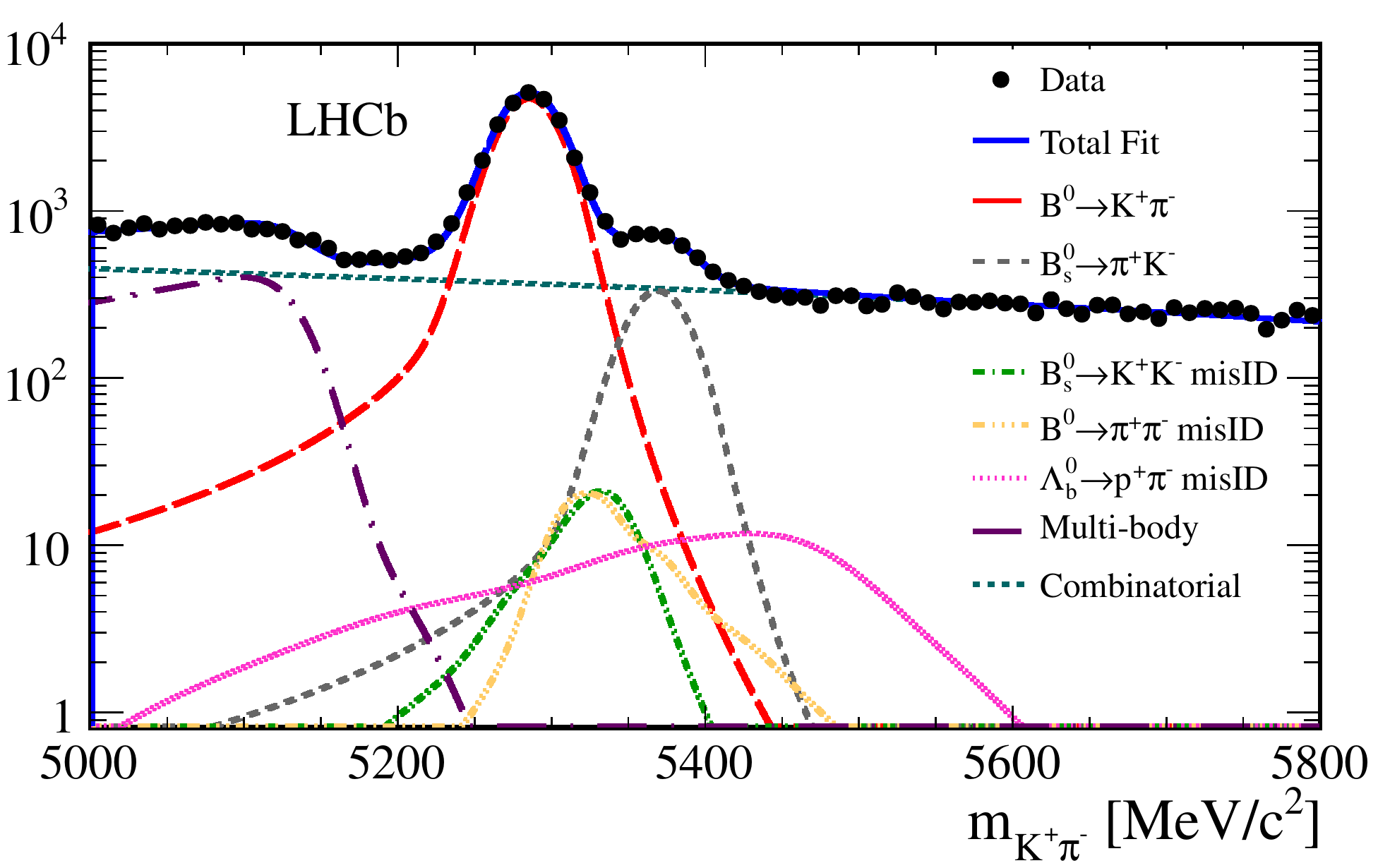}
     \includegraphics*[width=0.49\textwidth]{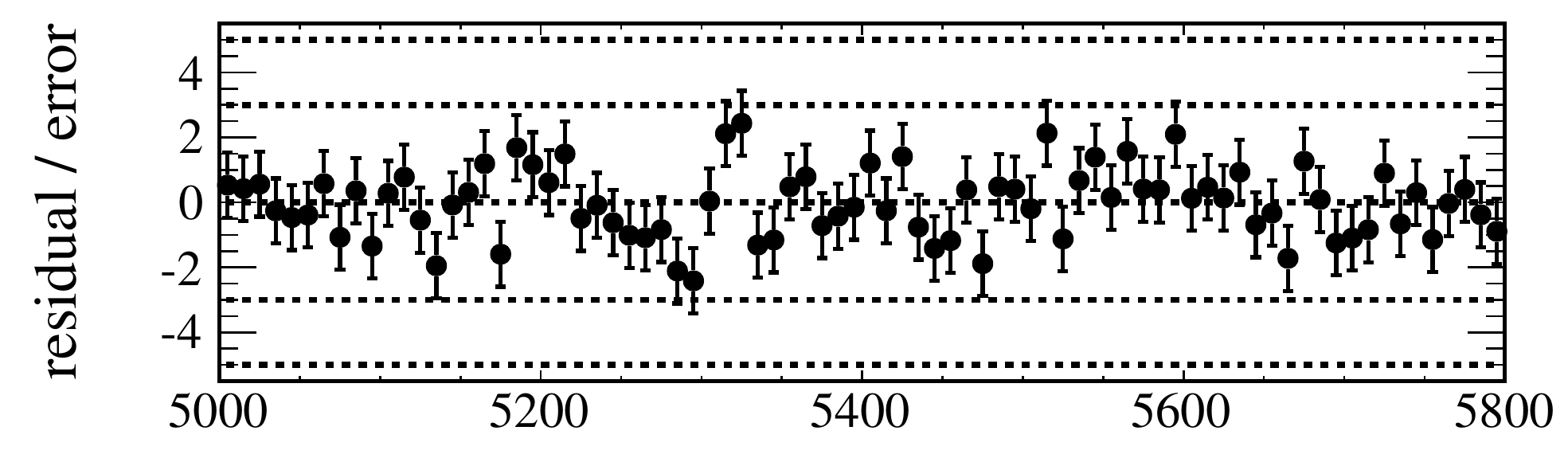}
     \includegraphics*[width=0.49\textwidth]{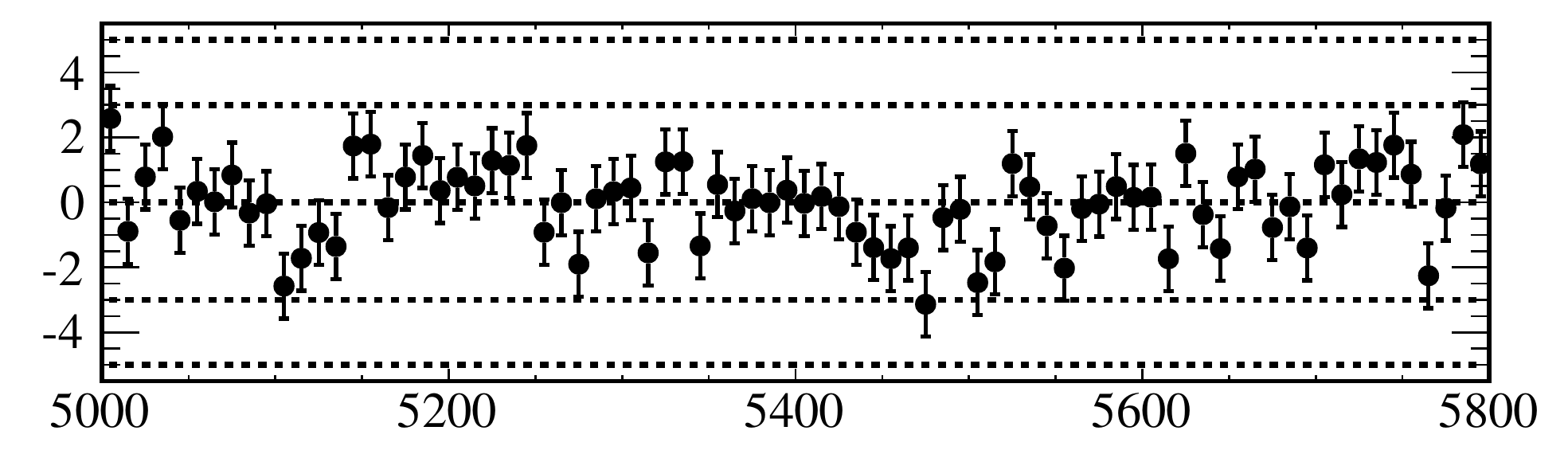}
  \end{center}
   \caption[\BTohh mass fit.]{\small Fits to the (left) $K^{+}K^{-}$ and (right) $K^{+}\pi^{-}$ invariant mass spectrum, with the main contributory signal and background
                                                             components displayed. The fit residuals are provided beneath each respective mass spectra.
					               }
  \label{fig:massFitBTohh}
\end{figure}

The reconstructed \BTohh mass and lifetime spectra include many contributions in addition to the combinatorial background, which arises from random combinations of reconstructed tracks. These backgrounds must be modelled accurately to reduce potential biases in the final measurement. The additional backgrounds consist of misreconstructed multi-body decays and misidentified physics backgrounds. Multi-body decays, such as the process $\Bd\rightarrow K^{+} \pim \piz$, may be reconstructed incorrectly as two body decays and in general populate a region of lower values than the signal in the mass spectrum. Misidentified backgrounds may originate from other \BTohh decays due to misidentification of the final state particles, where the correctly identified \BdToKK is treated as a misidentified background in the fit to the $K^{+}K^{-}$ spectrum. The \BdToKK decay is treated this way due to its relative contribution being too small to fit for using a parametrised function.
\newline
\indent The effective \BTohh lifetimes are extracted using an unbinned maximum likelihood fit in which probability density functions (PDFs) are used to describe the mass and decay time distributions. The measurement is performed by factorising the process into two independent fits, where the mass and decay time have been verified to be uncorrelated by correlation plots and comparing the decay time distribution in different mass intervals for the combinatorial background. The first fit is performed to the observed mass spectrum, see Fig.~\ref{fig:massFitBTohh}, and is used to determine the signal and background probabilities of each candidate. The yield of each misidentified background is fixed to the yield of the primary signal peak, \BsToKK or \BdToKpi, using the world average branching fractions and measured hadronisation ratios~\cite{bib:LHCb_fsfd}. The deviation visible around the \BdToKK peak in the \BsToKK mass fit, Fig.~\ref{fig:massFitBTohh}~(left), may be due to limited knowledge of the \BdToKK branching fraction. The mass fit probability density $f(m)$ can be written as the sum over the individual PDFs, $f(m|\mathrm{class})$, for all signal and background classes multiplied by the corresponding relative yield of that class $P(\mathrm{class})$, 

\begin{equation}
  f(m)=\sum_{\mathrm{class}}f(m|\mathrm{class})\cdot P(\mathrm{class}),
  \label{eqn:fit_mass_general}
\end{equation}

\noindent where $m$ is the measured mass of the candidate. The PDF models used to describe the mass distributions of each class are determined from full \lhcb simulation, with the exception of the multi-body background in the $K^{+}K^{-}$ spectrum that uses both simulation and data for its description. A sum of two Crystal Ball (CB) functions~\cite{bib:CBDist}
describes the \BsToKK, \BdToKpi and \BsTopiK signal decays. Misidentified background classes are described by template models extracted from simulation. The multi-body background is described using an exponentially modified Gaussian distribution,\footnote{$f(m;\mu,\sigma,\lambda) = \frac{\lambda}{2}e^{\frac{\lambda}{2}(2\mu+\lambda\sigma^{2}-2m)}\mathrm{erfc}(\frac{\mu+\lambda\sigma^{2}-m}{\sqrt{2}\sigma})$} while the combinatorial background component is modelled with a first order polynomial. Only candidates in the mass range $5000-5800$ \mevcc are used, with 22\ 498 and 60\ 596 candidates contributing to the $K^{+}K^{-}$ and $K^{+}\pi^{-}$ spectrum, respectively. The fits to each invariant mass spectrum yield $10\ 471 \pm 121$ \BsToKK, $26\ 220 \pm 200$ \BdToKpi and $1891\pm85$ \BsTopiK signal events.
In addition, the \sWeights~\cite{bib:sPlot}, signal fractions $P(\mathrm{class})$ and the probability of an event belonging to a particular signal class are also calculated by the mass fit and are used in the subsequent lifetime fit.

A fit to the reconstructed decay time spectrum is performed to measure the effective lifetime. The spectrum is described by a single exponential function, using a per-event acceptance correction calculated from data. The method used to evaluate the acceptance correction is detailed in Refs.~\cite{bib:AGamma2011, bib:Glasgow_LHCb_BsKKLifetime_2010_Paper}. The per-event acceptance functions are determined by moving each primary vertex along the momentum vector of the corresponding \B particle, and re-evaluating the selection for each emulated decay time. This procedure is repeated for a large number of hypothetical PV positions to verify whether a candidate would have been selected at that decay time. The set of decay times at which the per-event acceptance function turns on and off is denoted by
$A$ in Eq.~\ref{eqn:fit_lifetime_general}. The decay time PDF is modelled using a description of the unbiased distribution multiplied by the per-event acceptance function, denoted by $f(t|A,\mathrm{class})$. The likelihood function per candidate is given by

\begin{equation}
  f(t,A|m) = \sum_{\mathrm{classes}} f(t|A,\mathrm{class}) \cdot
  f(A|\mathrm{class}) \cdot \frac{P(\mathrm{class})f(m|\mathrm{class})}{f(m)},
  \label{eqn:fit_lifetime_general}
\end{equation}

\noindent where $t$ is the reconstructed decay time and $f(A|\mathrm{class})$
is the observed distribution of $A$ determined by the \emph{sPlot}
technique. The last factor is the probability for the candidate to
belong to a particular signal class.

The decay time PDFs of the background classes are modelled differently
from the signal. The misidentified \BTohh backgrounds are described
using an exponential function, with each lifetime fixed to the
respective current world average~\cite{PDG2012}. 
This is an approximation as these decays are reconstructed under the 
wrong mass hypothesis and a systematic uncertainty is 
assigned in Sect.~\ref{sec:systematics}.
The decay time PDFs of both the
multi-body and combinatorial background are estimated from data using
a non-parametric method involving the sum of kernel
functions~\cite{bib:Cranmer}. These functions represent each candidate
with a Gaussian function centred at the measured decay
time, with a width related to an estimate of density of
candidates at this decay time~\cite{bib:Cranmer} and normalised by
the \sWeight~\cite{bib:sPlot} of the candidate. The density of candidates
is estimated by the \sPlot~\cite{Pivk:2004ty} of the decay time distribution for each signal class.

This procedure approximates the observed decay time distribution,
including the acceptance effects. The fit method requires unbiased
decay time distributions since these are multiplied by the per-event
acceptance functions. The unbiased distributions are calculated
from the estimated observed distribution divided by the average
acceptance functions. The average acceptance function is calculated 
from an appropriately weighted sum of the per-event acceptance functions.

\begin{figure}[t!!!!]
  \begin{center}
    \includegraphics*[width=0.49\textwidth]{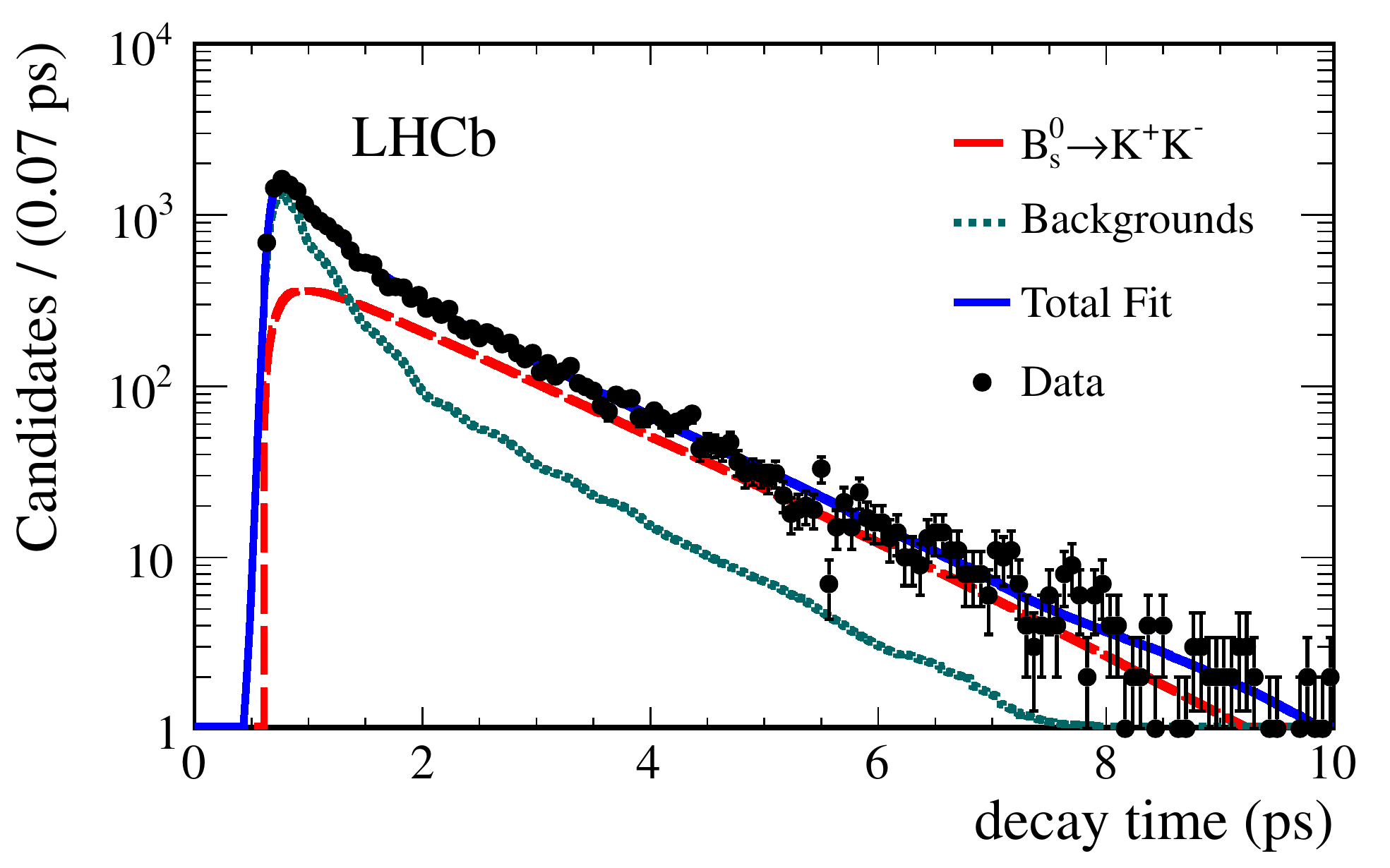}
    \includegraphics*[width=0.49\textwidth]{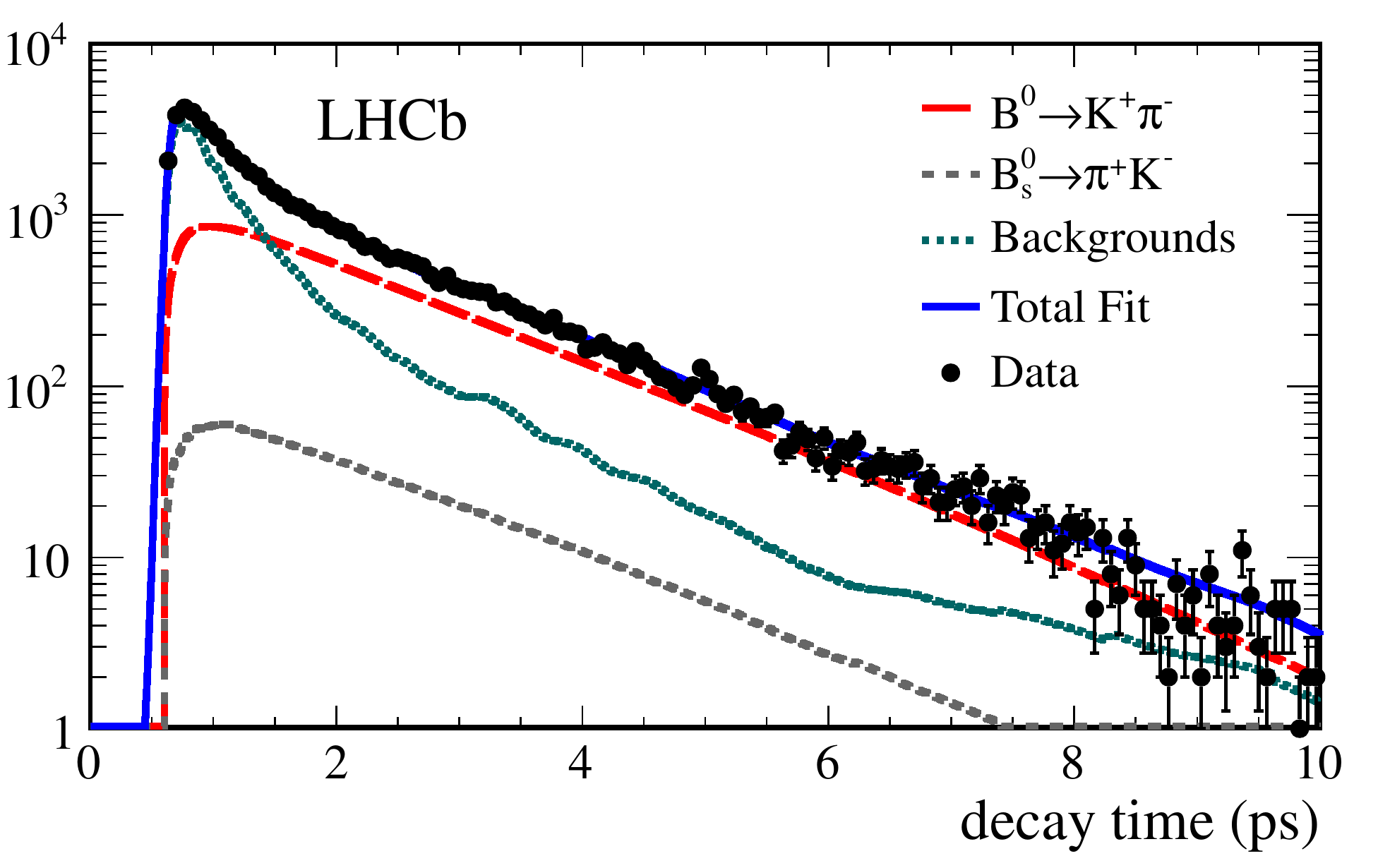}
    \includegraphics*[width=0.49\textwidth]{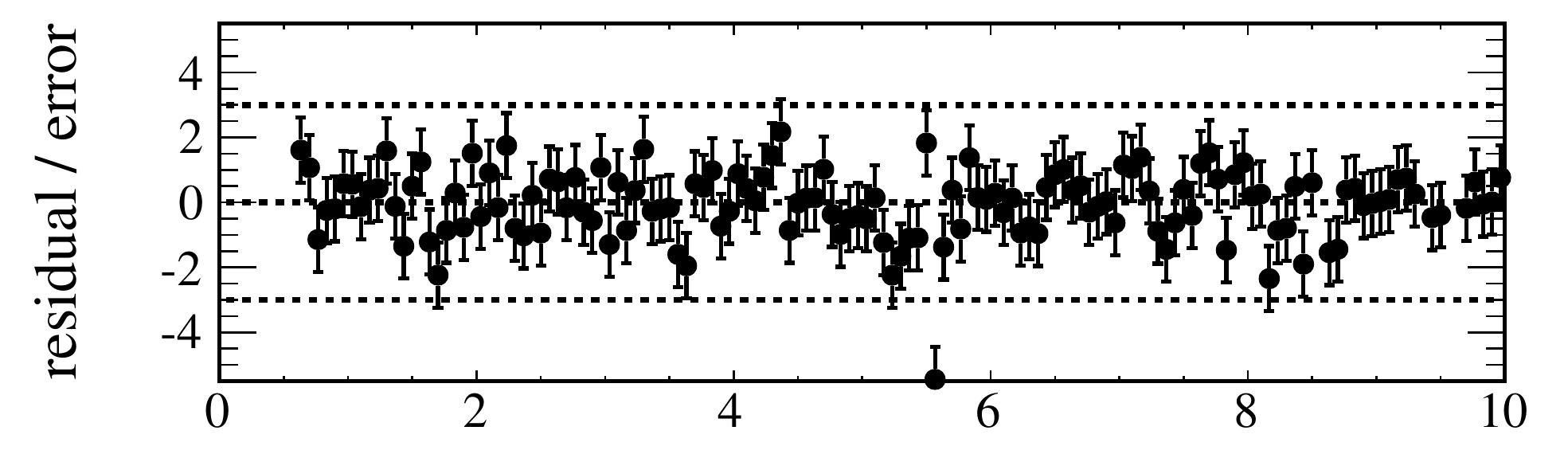}
    \includegraphics*[width=0.49\textwidth]{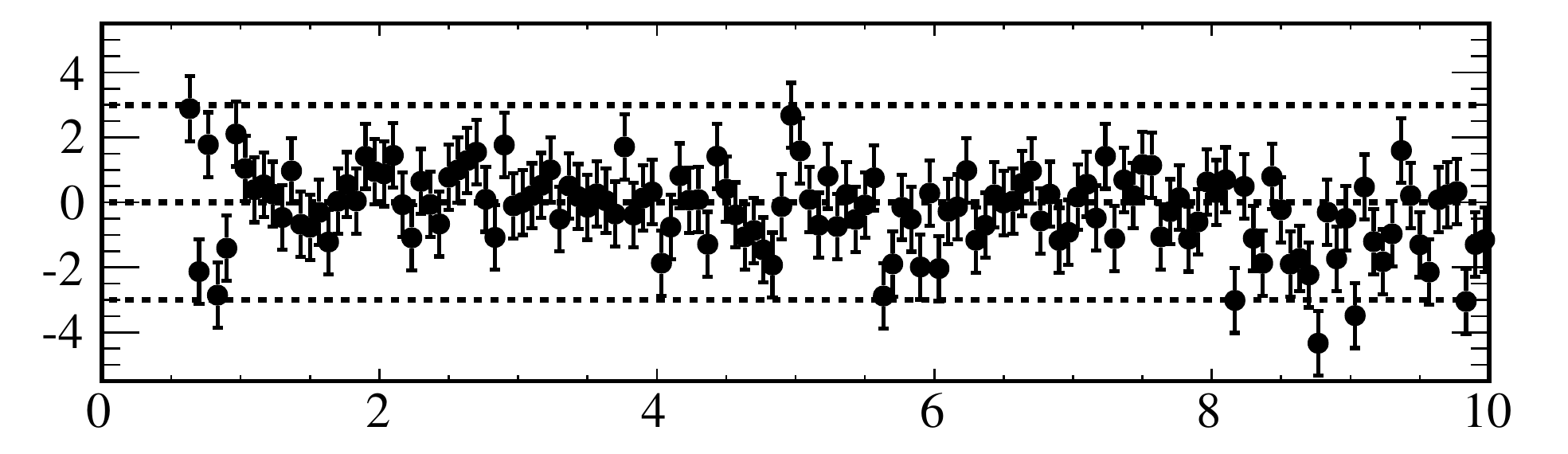}
  \end{center}
  \caption[\BTohh lifetime fits.]{\small Fit to the reconstructed
    decay times of the (left) \BsToKK decay and simultaneous fit to
    the (right) \BdToKpi and \BsTopiK reconstructed decay times.  The
    background distribution is the sum of all backgrounds displayed in
    Fig.~\ref{fig:massFitBTohh}. The fit residuals are provided beneath each respective decay time spectra.}
  \label{fig:LTFitBTohh}
\end{figure}

The lifetime fit is performed in the decay-time range $0.61-10.00\ps$,
due to a decay time cut of $0.60\ps$ in the selection
and to ensure that a sufficiently large number of candidates is
available for the method to be stable.
The fit results for the \BTohh channels are displayed in
Fig.~\ref{fig:LTFitBTohh}.

\section{Systematic Studies}
\label{sec:systematics}

The systematic uncertainties are listed in Table~\ref{tab:syst} and
discussed below. 

\begin{table}
  \begin{center}
    \caption{\label{tab:absSyst} Systematic uncertainties on
       the effective lifetimes. The uncertainties vary between the \BsToKK,
      \BdToKpi and \BsTopiK measurements due to the available
      sample size per decay mode.}
    \begin{center}\begin{tabular}{lccc}
      \hline
      Source & \multicolumn{3}{c}{Uncertainty (fs)} \\ 
             & \BsToKK & \BdToKpi & \BsTopiK \\
      \hline
            
      Cross contamination                 &  4.8   &  1.3   &  6.0    \\
      Tracking efficiency                 & 2.8 & 2.8 & 2.8 \\
      Mass model                          & 1.1 & 2.5 & 6.7 \\
      $B_{c}^{+}$ contamination            & 1.1 & -- & 1.1 \\
      Non-parametric decay time modelling  & 0.8 & 1.6 & 6.7 \\
      Production asymmetry                & 3.0 &  --   &  --   \\
      Effective lifetime interpretation   & 1.2 & --  & --  \\
      Remaining uncertainties & 0.8 & 0.6 & 3.7\\
      
      \hline
      Total                               & 6.7 & 4.3 & 12.2 \\
      \hline
    \end{tabular}\end{center}
    \label{tab:syst}
  \end{center}
\end{table}

The dominant contribution to the systematic uncertainty, in particular
for the \BsToKK and \BsTopiK effective lifetimes, comes from the
contamination from misidentified \BTohh background channels.
To determine the relative contribution of the most significant
misidentified backgrounds, we first determine the misidentification
probability of protons, pions and kaons as measured in data using the
decays $\KS \ra \pip \pim$, $\Dz \ra \Km \pip$, $\phi \ra \Kp \Km$ and
$\Lambda \ra p \pim$, where the particle type is deduced without using
PID information.
The particle identification likelihood method used to
separate pions, kaons and protons  depends on kinematic and global
event information such as momentum, transverse momentum, and the number of
reconstructed primary interaction vertices. The events in the calibration samples are weighted to match the
distributions of these variables in the signal sample.
The mass spectrum of the misidentified backgrounds are fitted
under the correct mass hypothesis to extract the yields, before being translated  
into cross-contamination rates using the PID efficiencies and misidentification rates. 
For the
sub-dominant backgrounds, the known branching and hadronisation
fractions are used to estimate the yields instead of the
fitted values. The value of the systematic uncertainty is given by
 the change in the fitted lifetime when the contamination rates are
varied within their uncertainty.

Another systematic uncertainty arises from the track reconstruction
efficiency and applies equally to all three decays. 
The track finding algorithm prefers tracks originating
from the beam-line, so those from long-lived \PB
decays have a slightly lower reconstruction efficiency. To determine
the impact of this uncertainty, the track reconstruction efficiency is
parametrised from data and then emulated in a large number of
simulated pseudoexperiments. Further details about this effect, and
its parametrisation, are provided in Ref.~\cite{bib:BetaFunctionRef}.
The difference between the generated and fitted lifetimes is
determined and the full offset is subtracted from the final fitted
lifetime and $50\%$ of the value is assigned as a systematic uncertainty.

The sensitivity to the details of the implemented signal and
background mass models are studied by varying the model parameters
taken from simulation. This systematic uncertainty
particularly affects the effective lifetime in the \BsTopiK decay. The
tail parameters of the double Crystal Ball function describing the signal peaks
and the parameters of the exponentially modified Gaussian function describing
the multi-body backgrounds are varied to accommodate the differences
between simulations and data.

The position of the mass shapes of the misidentified backgrounds are
fixed relative to the position of the signal peaks. The offset
is varied from the central value by the uncertainty of the 
mean of the fitted primary signal peak to 
determine the effect on the fitted lifetime.

The sensitivity to the shape of the combinatorial background
model is estimated by changing the description from a first-order
polynomial to an exponential function, 
the uncertainty being given by the lifetime difference observed.

The effective lifetimes in the \BsToKK and \BsTopiK decays are also
affected by contamination from secondary \Bs mesons decaying from
$\Bc$ mesons. Studies of the $\B_{c}^{+}\rightarrow \Bs\pip$ decay give
an upper limit of $1\%$ on the fraction of \Bs mesons
that originate from \Bc decays~\cite{bib:BcToBspi_2013_Note}. The
systematic uncertainty is estimated from simulated pseudoexperiments,
by adding a lifetime contribution that represents the \Bc decays
with the resultant deviation from the expected lifetime
assigned as the uncertainty.

The sensitivity to the modelling of the non-parametric component of
the background, which comprises the multi-body and combinatorial
backgrounds, is tested using three approaches. The first is estimated
by varying the width of the Gaussian kernels~\cite{bib:Cranmer} to determine
their effect on the background decay time distributions. The second is
studied by varying the decay time depending on the mass of the combinatorial
background. This is performed by splitting the mass range
$5480 - 5880\mevcc$ into three bins and ensuring that the decay time
distribution in each bin shows no variation within its statistical
uncertainty. The final study estimates the systematic uncertainty
assuming a correlation between the decay time distribution and the
mass of the multi-body and misidentified backgrounds. 
This is performed using simulated pseudoexperiments, where the generated 
lifetime is scaled by a factor determined by the ratio of the 
generated misreconstructed mass and the true mass for each event.
The modelling of the non-parametric background has the largest influence on
the effective lifetime in the \BsTopiK decay since the signal significance is the
smallest in this channel.

The analysis assumes that \Bs and \Bsb mesons are produced in equal
quantities. Deviations from this assumption affect the effective
lifetime in the \BsToKK decay but not the effective lifetimes in the
flavour specific decays. The production asymmetry is measured
experimentally to be $(7\pm5)\%$~\cite{bib:ProductionAsymmetry2013},
and its influence is evaluated from an analytical calculation
using the current experimental values.

This analysis presents a measurement of the effective \BsToKK
lifetime, which is equivalent to measuring the decay time using a
single exponential function and is commonly evaluated using the
formula described in Ref.~\cite{bib:Hartkorn99}. This is only valid in
the absence of acceptance effects. The a priori unknown fractional
components of the light and heavy mass eigenstates that contribute to
the decay of the \Bs meson result in an interpretation bias. Using
conservative choices for \DGs and \ADGs, the size of the effect is
studied with simulated pseudoexperiments. The result is labelled
``Effective lifetime interpretation'' in Table~\ref{tab:syst} and is
treated as a source of systematic uncertainty on the measurement.

The remaining sources of uncertainty are the following: the precision
at which the fitting method was verified; the uncertainty on the world average
lifetimes used to model the misidentified \BTohh backgrounds; the
modelling of the decay time resolution in the lifetime fit; the
absolute lifetime scale given by the alignment and the absolute length
of the VELO. These are all individually small and sum
up to the last line in Table~\ref{tab:syst}.

The method itself is verified as being unbiased using simulation of the \lhcb
experiment and a large number of simulated pseudoexperiments.

Additionally, studies of the effect of the trigger, primary vertices
and magnet polarity are performed. The data are divided into
subsets corresponding to periods with different magnet
polarity, trigger configuration, and for different numbers of primary
vertices. These have no effect on the measured lifetime and therefore
no systematic uncertainty is assigned.

\section{Results and conclusions}
\label{sec:conclusions}

The effective \BsToKK lifetime is measured in $pp$ interactions using
a data sample corresponding to an integrated luminosity of $1.0\invfb$
recorded by the \lhcb experiment in 2011. A data-driven approach is used to
correct for acceptance effects introduced by the trigger and final
event selection. The measurement evaluates the per-event
acceptance function directly from the data and determines the
effective lifetime to be
\begin{equation*}
  \tau_{\BsToKK} = 1.407\pm0.016\stat\pm0.007\syst\ps,
  \label{eq:Bs2KKLifetime}
\end{equation*}

\noindent which is compatible with the prediction of $1.395\pm0.020\ps$. The measured value is significantly more precise than and
supersedes the previous \lhcb measurement of this effective lifetime from the same dataset~\cite{bib:lhcbBs2KKUnbiased}, but is statistically independent of the result in Ref.~\cite{bib:Glasgow_LHCb_BsKKLifetime_2010_Paper}. This measurement can be combined with measurements of \DGs and \Gs, given in Ref.~\cite{bib:LHCB-PAPER-2013-002}, to make a first direct determination of the asymmetry parameter \ADGs to first order using 
\begin{equation}
\label{eq:ADeltaGammaEq}
\ADGs = \frac{2\Gamma_{s}^2}{\Delta\Gamma_{s}}\tau_{\BsToKK} -  \frac{2\Gamma_{s}}{\Delta\Gamma_{s}}.
\end{equation}

\noindent The value is found to be
\begin{equation*}
  \ADGs = -0.87\pm0.17\stat\pm0.13\syst,
  \label{eq:ADeltaGammaCalc}
\end{equation*}

\noindent which is consistent with the level of \CP violation predicted by the SM~\cite{bib:Fleischer1}. In the limit of no \CP violation, the effective \BsToKK lifetime corresponds to a measurement of $\Gamma_{\mathrm{L}}$ of
\begin{equation*}
  \Gamma_{\mathrm{L}} = 0.711\pm0.008\stat\pm0.004\syst\ps^{-1}.
  \label{eq:GammaLMeas}
\end{equation*}

\noindent This is compatible with the value of $\Gamma_{\mathrm{L}}$ determined from the $\Bs\rightarrow D^{+}_{s}D^{-}_{s}$ channel in Ref.~\cite{bib:PhysRevLett.112.111802}. In addition, measurements of the effective \BdToKpi and \BsTopiK lifetimes are also performed with the same method. The measured effective lifetimes are
\begin{equation*}
  \tau_{\BdToKpi} = 1.524\pm0.011\stat\pm0.004\syst\ps,
  \label{eq:Bd2KpiLifetime}
\end{equation*}
\begin{equation*}
  \tau_{\BsTopiK} = 1.60\pm0.06\stat\pm0.01\syst\ps.
  \label{eq:Bs2piKLifetime}
\end{equation*}

The measured \Bd effective lifetime is compatible with the current world average of $1.519\pm0.007\ps$\cite{PDG2012}, with the effective lifetime of the flavour-specific \Bs compatible within $2\sigma$ of its respective world average of $1.463\pm0.032\ps$~\cite{PDG2012}.




\section*{Acknowledgements}

 
\noindent We express our gratitude to our colleagues in the CERN
accelerator departments for the excellent performance of the LHC. We
thank the technical and administrative staff at the LHCb
institutes. We acknowledge support from CERN and from the national
agencies: CAPES, CNPq, FAPERJ and FINEP (Brazil); NSFC (China);
CNRS/IN2P3 (France); BMBF, DFG, HGF and MPG (Germany); SFI (Ireland); INFN (Italy); 
FOM and NWO (The Netherlands); MNiSW and NCN (Poland); MEN/IFA (Romania); 
MinES and FANO (Russia); MinECo (Spain); SNSF and SER (Switzerland); 
NASU (Ukraine); STFC (United Kingdom); NSF (USA).
The Tier1 computing centres are supported by IN2P3 (France), KIT and BMBF 
(Germany), INFN (Italy), NWO and SURF (The Netherlands), PIC (Spain), GridPP 
(United Kingdom).
We are indebted to the communities behind the multiple open 
source software packages on which we depend. We are also thankful for the 
computing resources and the access to software R\&D tools provided by Yandex LLC (Russia).
Individual groups or members have received support from 
EPLANET, Marie Sk\l{}odowska-Curie Actions and ERC (European Union), 
Conseil g\'{e}n\'{e}ral de Haute-Savoie, Labex ENIGMASS and OCEVU, 
R\'{e}gion Auvergne (France), RFBR (Russia), XuntaGal and GENCAT (Spain), Royal Society and Royal
Commission for the Exhibition of 1851 (United Kingdom).



\addcontentsline{toc}{section}{References}
\setboolean{inbibliography}{true}
\bibliographystyle{LHCb}
\bibliography{main,LHCb-PAPER}

\end{document}